# COLLECTIVE ACCELERATION OF IONS BY MEANS OF PLASMOIDS IN RF WELLS OF FREQUENCY-MODULATED LASER FIELD.

A.I. Dzergatch and S.V. Vinogradov

Moscow Radiotechnical Institute, 113519 Moscow, Russia

## 1. INTRODUCTION

The proposed linear accelerator ("scanator") consists of a terawatt table-top laser and a set of passive elements - beam splitters, dispersion elements for stretching of the laser pulse and chirping of the splitted beams, and dispersion elements for angle scanning of crossed frequency-modulated laser beams. Ions are trapped and accelerated in RF wells by the electron component of plasmoids in the intersection zone of the scanning laser beams. Computational studies give encouraging results. A proof-of-principle experiment on the base of a table-top laser is outlined.

Several groups investigate accelerators based on plasma waves, which are excited by powerful short laser or electron pulses (look, *e.g.*,[1] and references therein).. These schemes are based on ***free*** oscillations of the plasma and hence they directly depend on the plasma tolerances and instabilities.

The present variant of acceleration is based on ***forced*** oscillations of the charged plasma in laser-generated moving or standing RF wells (HF traps, ponderomotive- or quasipotential wells, M.A. Mil-ler's force, light pressure. This way leads to several schemes of regular acceleration, based on far fields. The dependence on plasma parameters is decreased in this variant. One of these schemes [2], namely MWA (moving well accelerator), is detailed and discussed in this report.

Certain vacuum modes of fast electromagnetic waves (far field) trap charged particles, electrons (positrons) in the 1-st turn, near the minimums of the envelope or near the zeros of the carrier frequency [2]. Both types of RFwells ("envelope wells" and "carrier wells") may be distant from the radiating surfaces, hence the electric breakdown problems are moved aside and concentrated fields with very high amplitudes may be used. The RF wells may be effective (gradient of the quasi-potential ~tens % of the field amplitude $E_m$), if the amplitude is large, $E_m \sim \mathrm{mc}^2 / el$ , *e.g.*, 1 TV/m in case of electrons and a 1-$m$ m laser.

This effect may be treated as 3-dimensional alternating gradient focussing of the electron component of plasmoids. The computed dimensions of plasmoids in case of carrier RF wells are ~λ/6 or smaller, and their den-sity is sub-critical, so they are not larger than several Debye lengths. It simplifies the plasma stability problems.

Motion and acceleration of an RF well takes place, if the given structure of its field in the moving frames (e.g., a cylindrical wave $E_{0mn}(j, r, z)$) is generated by corresponding laboratory sources (the moving and laboratory fields are connected by Lorentz transforms).

## 2. THE STRUCTURE OF THE FIELD AND THE SCHEME OF THE ACCELERATOR

The field structure is based on A.M. Sessler's idea [3] to use crossed beams of a small laser instead of the expensive system of oversized resonators with kilojoules of stored optical energy. The RF wells exist in many points in the zone of intersection of the focused laser beams.

These beams are crossed and focused (Fig.1) at the center P of an RF well, which is accelerated along the z-axis, if the field parameters have the proper variations. The programs of the frequencies and angles variations are defined by Lorentz-trans-formed values of the RF well parameters, $w, q$ , prescribed in the moving frames. The frequency $w$ may be constant, but the inevitable variation of the angle $q$ , *i.e.*, of the RF well form, limits its values near $45 \pm 15^o$ . The sources of these beams (focused dispersion radiators at the Fig.2) are centered at 8 points ($\pm x_1$; 0), (0;$\pm y_1$), ($\pm x_2$; 0) and (0; $\pm y_2$), symmetric in the planes *xz* and yz. The number of these partial beams may vary, in principle, between 6 and infinity (cylindrical waves).

During the acceleration the beams are scanned (from left to right at the Fig.1). This process is realized by linear transforms (filtering) of the primary short (wide-band) pulse of the feeding laser. This pulse is split into a pair of pulses, and each of them is stretched and frequency-modulated (FM, chirped) by means of, positive and negative dispersion elements ±D.

The lags of the ion center from the electron center and of the latter – from the RF well center must be small (say, 0.01 $l$ ), if the number of accelerated ions must be large; its increase leads to a decrease of the number of accelerated ions. Some excess of electrons ensures the longitudinal autofocalization of ions.

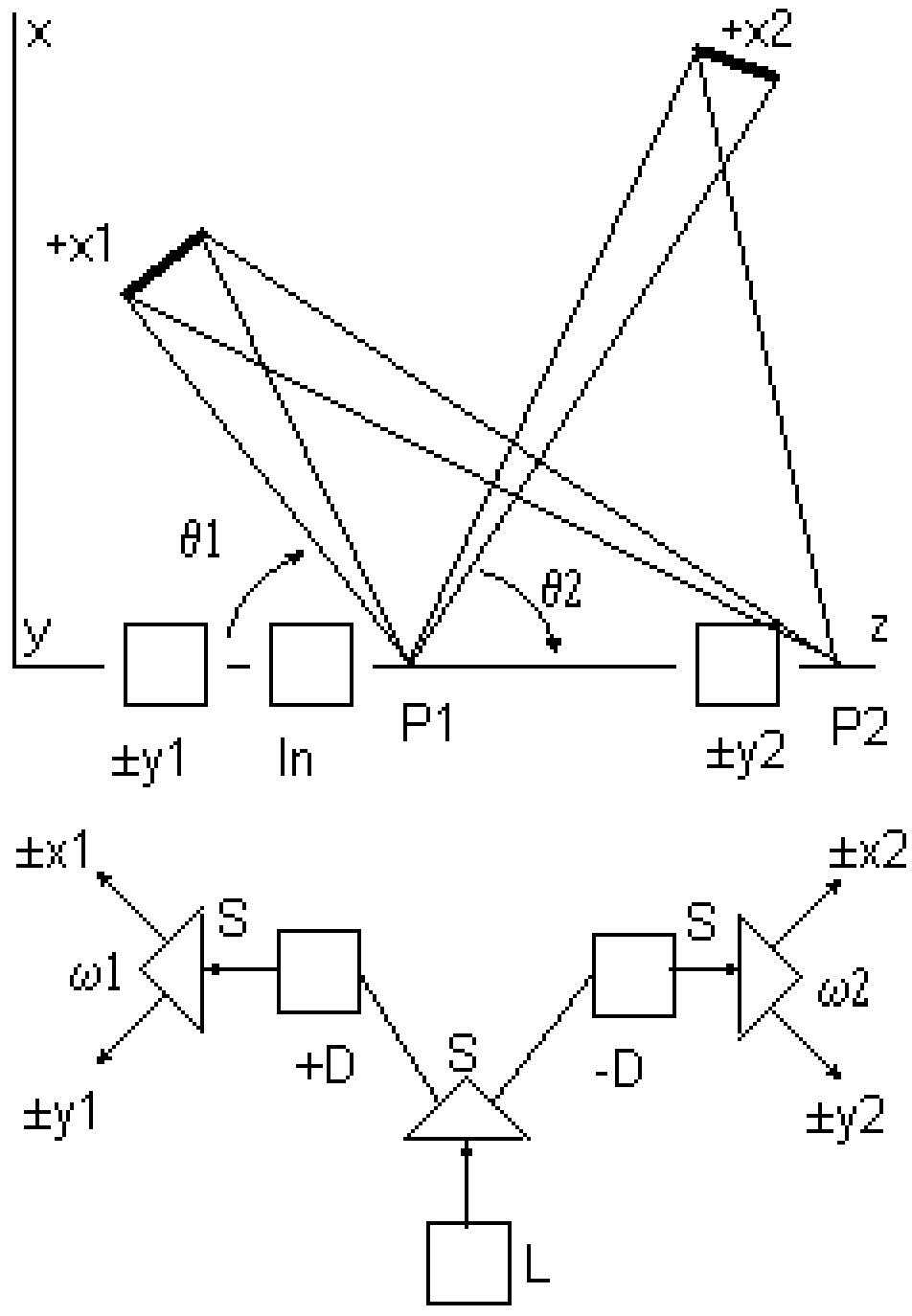

Fig 1. Scheme of the scanator.

The injector may be simply a gas jet similar to that used in printers. Some additional radiators (not shown) may be installed (and fed from the same laser) as correctors, if needed.

Estimated parameters of a proof-of- principle model proton accelerator (Fig.1) are given in the Table 1 below:

Table 1. Some parameters of the scanator model.

| | |
|---|---|
| Laser pulse energy/peak power: | 3 J/300 GW |
| Laser wavelength | ~1 mkm |
| Diameter of the FDR: | 7 mm |
| Length of the acceleration path: | 5 cm |
| Distance radiator-acceleration zone: | 15 cm |
| Maximal angle scans: | ~1 grad |
| FM deviations: | $\pm$ 1 % |
| Number of RF wells in the focal region: | ~ 500 |
| Focal field density | 200 GV/m |
| Neutralization factor | ~0.8 |

The number of accelerated ions per plasmoid is defined by the ion density and by the plasmoid volume, and it is proportional to the ratio $r_e / l$. The accelerated current does not depend on the wavelength $l$ (at a given relative density $n / n_c$).

The state of the art of tera- to peta-watt subpicosecond lasers gives hope on the realization of the proposed scheme.

The above variant of the "scanator" is based on the "carrier RF wells", which are disposed with z-intervals equal to a half of the z-wavelength. These wells are relatively small, which simplifies the plasma stability problems.

## 3. METHODS AND RESULTS OF NUMERICAL STUDIES

"Multi-particles" programs are used for finding the tolerable densities of the electron and ion components of plasmoids and for final checking of the acceleration concept in various regimes, including long computations (~50 000 RF periods).

The axially-symmetric relativistic motion of many electromagnetically interacting electrons and protons was modeled in the *rz*-plane by the PIC method (2.5 measurements, $r, r', z, z'$, and the full velocity *v*, rectangular toroidal macroparticles). Full system of Maxwell's equations and the equations of macroparticles motion were solved for electrons and ions in the co-moving (with the accelerated plasmoid) ideal cylindrical resonator tuned to the same wave $E_{011}$ as in the 1-particle case. The code was written in C++ language. The number of macroparticles in the calculations was usually ~50 000, the grid sizes about 30×30. The use of moving frames leads to large economy of computation time. Special checks (longitudinal waves in tubular beams, transverse waves in plasma columns, several modes in an empty cylindrical resonator) have shown the precision better than several %. So the use of this "computational" resonator is justified for the present case, when the plasmoid is relatively small, ~ $l$ /6 or less. The additional physical parameters in the multi-particle case are the initial densities of electrons and protons in the charged plasmoid and some computational parameters (the numbers of computation cells and steps per RF period, *etc*). Preliminary values of the densities were chosen with the account of Kapchinski — Vladimirski equilibrium and its stability studies, which lead to the AG focusing depression by the space charge up to ~30%.. So the initial conditions were uniform density and zero velocity for both electrons and ions, which lead to very non-uniform density and losses ~20% of the particles at the initial several hundreds of periods.

The physical parameters were field amplitude, Brillouin angle, initial acceleration, 2 densities. Some typical shots of electrons (upper bunches) and ions (lower bunches) are shown at the Fig-s 2-3 for the times 79977 and 650007 time units 2.2 *λ/33c.* The initial distribution of electrons and protons was chosen (for the economy of cells) as a spheroid, corresponding to the RF well dimensions found in the preliminary 1-particle modeling. Fig.4 shows the numbers of electrons and protons in the accelerated RF well as functions of time (in units 2.2λ/*c*): after an initial relatively swift (~1000 field periods ) loss the self-consistent evolution process leads to acceleration of the particles during ~50 000 periods with a relatively small loss. The form of both bunches, electron and proton, is gradually normalized, and then a slow "evaporation" of particles takes place. This process is similar to halo formation in the case of RFQ linac.

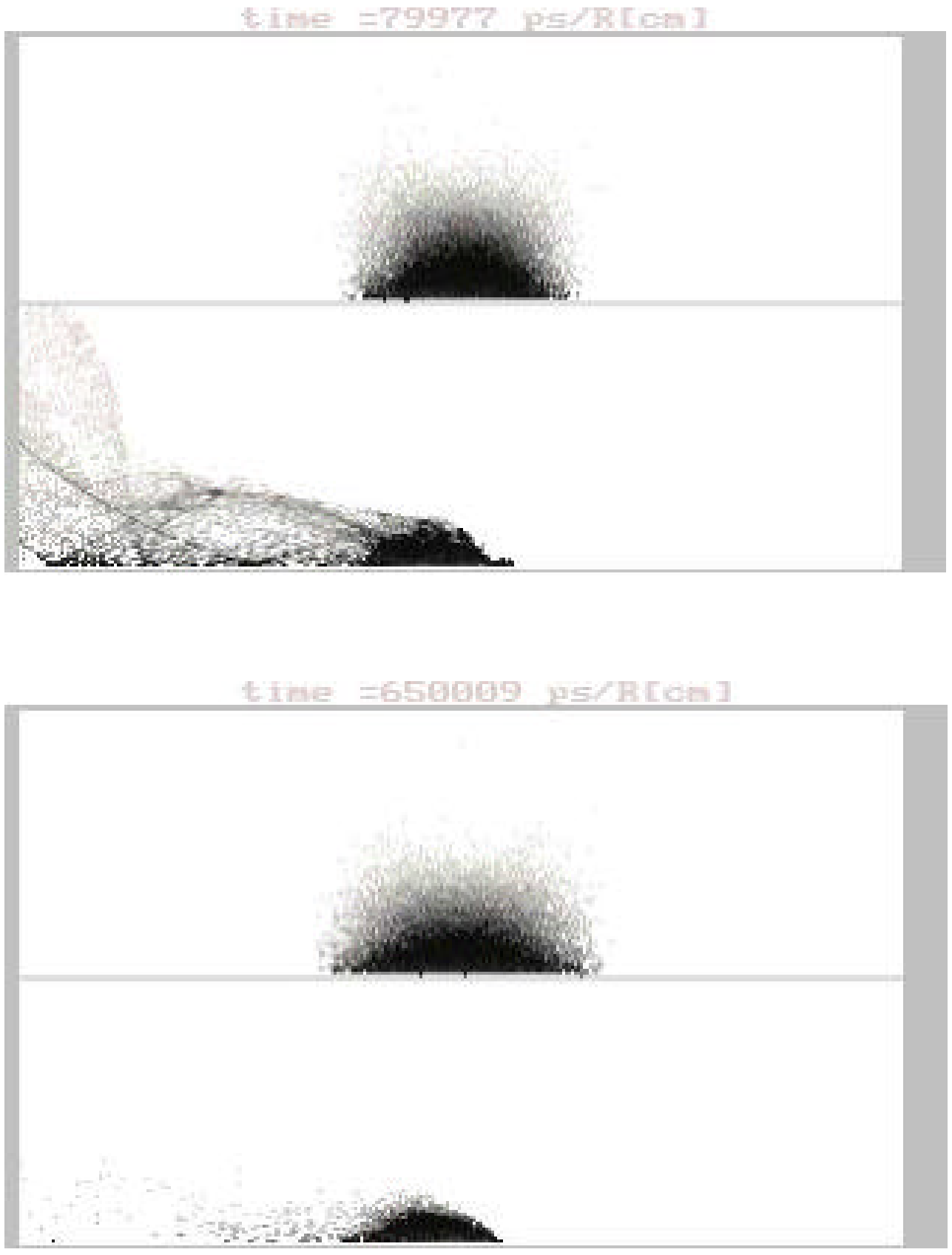

Fig 2-3. r-z portraits og electrons and accelerating ions in RF well

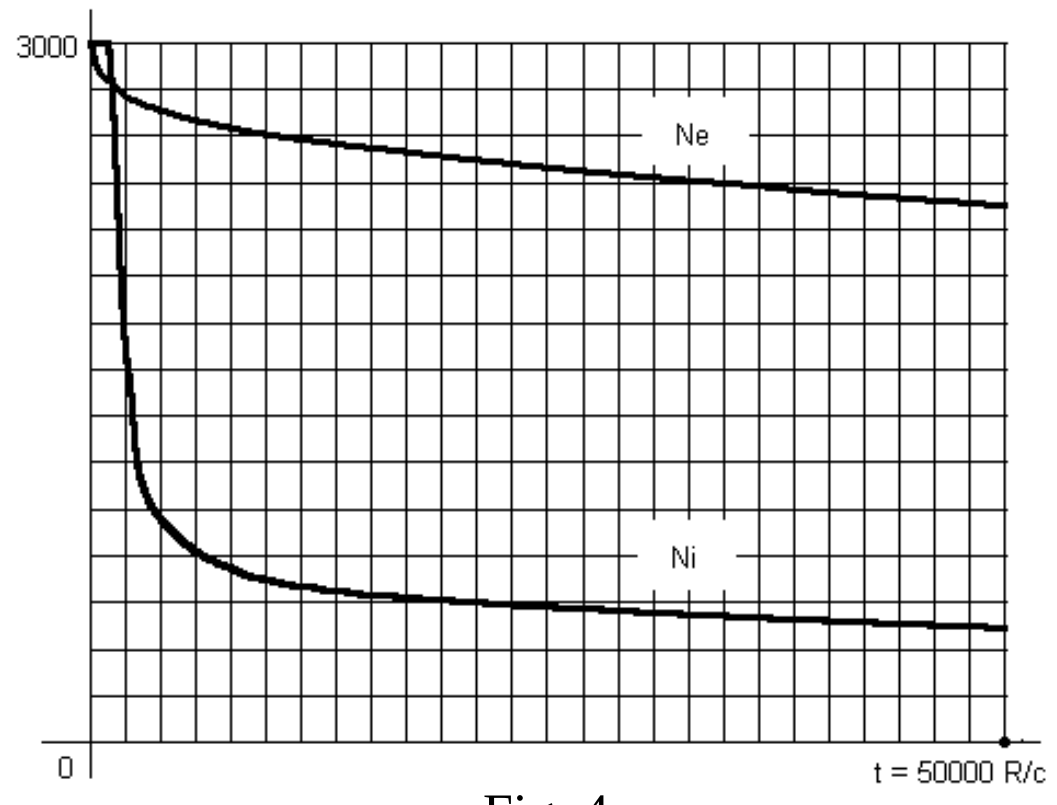

Fig. 4.

The computed cartoons show the alternating focusing-defocusing *rz*-oscillations, and the lag of accelerated ions from electrons, and of the electrons – from the RF well center. Optimal amplitude of the field was found to be $E_m \approx mc^2/R\,e$, where $R$ is the radius of the resonator, $R = 2.2\,\lambda$ for the present case of Brillouin angle $\theta \approx 60^\circ$, *e* is the electron charge.

The number of accelerated particles per plasmoid (which decreases with the increase of the acceleration) was found to be ~3000 electrons and ~1000 protons. The initial value of the acceleration (it decreases with growth of mass of the ions) was chosen in one of the computational runs to be 0.000 001 $c^2/R$, which corresponds to the acceleration gradient $dW/dz = Ma \approx 500$ MeV/m.

## CONCLUSION

A compact proof-of principle collective accelerator ("scanator") may be built on the base of a table-top terawatt laser and a passive optical sys-tem, which splits and transforms the primary laser beam into several frequency-modulated crossed scanning light beams. The ions are accelerated by the electron component of plasmoids (short plasma bunches), which are trapped by moving RF wells (HF traps) of the electromagnetic field in the in-tersection region. This region is periodically scan-ned along the line of acceleration. An estimation of parameters shows the possibility of acceleration of, say, protons, from a gas jet to 300 MeV, using a table-top terawatt 1-mkm laser and a set of usual optical elements (mirrors, prisms, diffraction gratings etc).

The numerical modeling confirms the possibility of collective acceleration by charged plasmoids in RF wells.

The encouraging computational results of the present studies show, amongst other, the desirability of theoretical nonlinear analysis of the problem which might lead to its better understanding.

## *ACKNOWLEGEMENTS*

This work was supported by the International Scientific and Technology Center. The authors are thankful to V.S.Kabanov, V.A.Kuzmin, and F.Amiranoff for the help and discussions.